\documentclass[11pt,twoside]{article}

\usepackage{asp2014}

\aspSuppressVolSlug
\resetcounters

\newcommand{\lstchain}{\texttt{lstchain} }
\newcommand{\ctapipe}{\texttt{ctapipe} }
\newcommand{\lstmcpipe}{\texttt{lstMCpipe} }

\begin{document}

\title{The lstMCpipe library}

\author{Enrique~Garc\'ia$^{1,2}$, Thomas~Vuillaume$^2$ and Lukas Nikel$^3$ }
\affil{$^1$IT Department, CERN, 1211 Geneva 23, Switzerland}
\affil{$^2$Univ. Savoie Mont-Blanc, LAPP, CNRS, Annecy, France}
\affil{$^3$TU Dortmund: Dortmund, Germany}

\paperauthor{Garc\'ia~E.}{enrique.garcia.garcia@cern.ch}{https://orcid.org/0000-0003-2224-4594}{CERN, IT Department}{}{Geneva}{}{1211}{Switzerland}
\paperauthor{Vuillaume~T.}{thomas.vuillaume@lapp.in2p3.fr}{https://orcid.org/0000-0002-5686-2078}{Univ. Savoie Mont-Blanc, LAPP, CNRS}{}{Annecy}{}{74370}{France}
\paperauthor{Nikel~L.}{}{https://orcid.org/0000-0001-7110-0533}{TU Dortmund}{}{Dortmund}{}{}{Germany}



\begin{abstract}
The Cherenkov Telescope Array (CTA) is the next generation of ground-based gamma-ray astronomy observatory that will improve the sensitivity of current generation instruments by one order of magnitude. The LST-1 is the first telescope prototype built on-site on the Canary Island of La Palma and has been taking data for a few years already.
Like all imaging atmospheric Cherenkov telescopes (IACTs), the LST-1 works by capturing the light produced by the Cherenkov process when high-energy particles enter the atmosphere. The analysis of the recorded snapshot of the camera allows to discriminate between gamma photons and hadrons, and to reconstruct the physical parameters of the selected photons. To build the models for the discrimination and reconstruction, as well as to estimate the telescope response (by simulating the atmospheric showers and the telescope optics and electronics), extensive Monte Carlo simulations have to be performed. These trained models are later used to analyse data from real observations.

lstMCpipe is an open source  python package developed to orchestrate the different stages of the analysis of the MC files on a computing facility. Currently, the library is in production status, scheduling the full pipeline in a SLURM cluster. It greatly simplifies the analysis workflow by adding a level of abstraction, allowing users to start the entire pipeline using a simple configuration file. Moreover, members of the LST collaboration can ask for  a new analysis to  be produced with their tuned parameters through a pull request in the project repository, allowing careful review by others collaborators and a central management of the productions, thus reducing human errors and optimising the usage of the computing resources.

\end{abstract}




\section{Introduction}

The Cherenkov Telescope Array (CTA) is the next generation observatory of Imaging Cherenkov Atmospheric Telescopes (IACTs) for ground-based gamma-ray astronomy \citep{cta2018}. It is now in the pre-construction phase but the first on-site prototype, the first Large-Sized Telescope (LST-1) is in the commissioning phase and already taking data on the Canary Island of La Palma (Spain) since 2018. 

\section{LST-1 data processing}\label{sec:cta_lst_processing}

To reconstruct the physical properties of the primary high energy particles that enters the atmosphere,  IACTs relies on Monte Carlo (MC) simulations. MC data is analysed to train machine learning models and to compute Instrument Response Functions (IRFs). LST-1 MC data are produced by the atmospheric shower generator CORSIKA \citep{1998cmcc.book.....H}, an open-source software chosen as the standard tool for CTA simulations and the simtel\_array package to simulate the telescopes optics and electronics responses \citep{ 2008APh....30..149B}.

MC and LST real data then follow the same reduction pipeline, from data level 0 (DL0, full waveforms data from telescopes) to data level 3 (DL3, reconstructed photons list) and Instrument Response Functions (IRFs). For LST-1, this data reduction can be done thanks to the \lstchain library \citep{lstchain22}\footnote{https://doi.org/10.5281/zenodo.7323874} which is based on the \ctapipe framework \citep{ctapipe-icrc-2021}. However, the orchestration of the different step for MC data and LST data analysis is very different and handled by two different libraries, \lstmcpipe for the MC data (this work) and LSTOSA \citep{lstosa} for the LST data.

\section{The lstMCpipe python library}

The \lstmcpipe library \citep{vuillaume_thomas_2022_7180216} is a Python package developed to orchestrate the MC data reduction pipeline steps implemented in \lstchain on the LST collaboration computing center. It takes advantage of the SLURM workload manager system \citep{slurm} to divide the analysis steps into jobs, handling their dependencies and the organization of their inputs and outputs in the file system.

\subsection{lstMCpipe workflow}

\begin{figure}
\includegraphics[width=\textwidth]{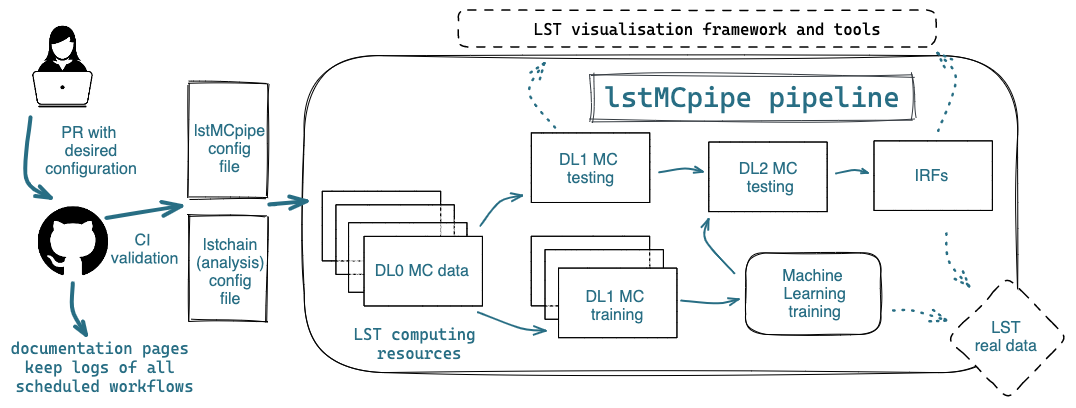}
\caption{A schematic view of the workflow handled by \lstmcpipe. First, the raw MC data DL0 is divided into training and testing and reduced to DL1 level, then merged per sub-datasets. The training dataset is used to the train RF models, which are then applied to the test dataset, that will then be used to derive the instrument response functions. RFs and IRFs are necessary products to analyse data from the LST-1.}
\label{fig:workflow}
\end{figure}

The MC data reduction pipeline is composed of the processing described in section~\ref{sec:cta_lst_processing} plus the training of the random forest (RF) models. From a set of MC data, four RF models are trained: a classifier for background rejection, and one classifier plus two regressors for the gamma-ray photon energy and incident direction reconstruction. These RFs will be used for LST real data processing, as well as to derive IRFs when applied to another MC dataset (see Figure~\ref{fig:workflow}).

\lstmcpipe automatically applies the different stage scripts (executable provided by \lstchain) to a full set of MC data, greatly simplifying the manual execution of a full pipeline. Each of the data reduction analysis steps implies executing a script on many files. For example, in the first stage of the MC data reduction (DL0 to DL1), the stage script can be applied to hundreds of thousands of simulated files. \lstmcpipe allows managing, parallelizing and orchestrating the execution of these set of scripts (or steps) in an ordered way, standardizing the intermediate and final outputs. An entire reduction pipeline can be (re)started from any stage in case of failure or of modification of an intermediate stage.

A typical \lstmcpipe workflow input is composed of 2 files: the \lstmcpipe configuration file describing the desired workflow, and a \lstchain configuration file that is passed along to the different analysis steps. The \lstmcpipe configuration file describes the stages to be run and their inputs and outputs, as well as the software environment to use. A typical output is composed of all the intermediate data level outputs (generated by the different \lstchain executable), the trained RF models and the IRFs. 

Because of the increasing complexity of the analysis, the library went through various updates that allowed managing the stages and the intermediate data level output in a more modular and flexible way. Currently, the package contains functionalities to automatically search the MC data in the LST cluster and generate a \lstmcpipe configuration file to be used for their analysis. This functionality makes the workflow setup transparent and user-friendly.

\subsection{Data analysis as a service}

In the LST-1 data-processing workflow, analyzers need to tune the MC data to their specific source in order to train and produce dedicated RFs and IRFs. In order to centralize these productions and minimize the errors and computing resources usage, we setup a production request service based on GitHub pull-requests. LST analyzers can thus open a simple pull-request (PR), including a \texttt{README.md} describing their request and why it is needed for their analysis, a \texttt{lstchain} configuration file and a \texttt{lstmcpipe} configuration file. This PR then goes through unit tests to validate the correctness of the configuration files and is reviewed by our team, checking for its validity.
Once accepted, the data production is launched manually on the cluster and the analyzer is notified when finished.
This centralization of the processes also allowed us to produce an online database in the project documentation web-page\footnote{https://cta-observatory.github.io/lstmcpipe/} with the logs of the different productions and the available trained models and IRFs. All members of the collaboration can then check if a set of model and IRFs fitting their need has already been produced.

\begin{figure}
    \noindent\begin{minipage}{0.5\textwidth}
        \centering
        \includegraphics[width=\textwidth]{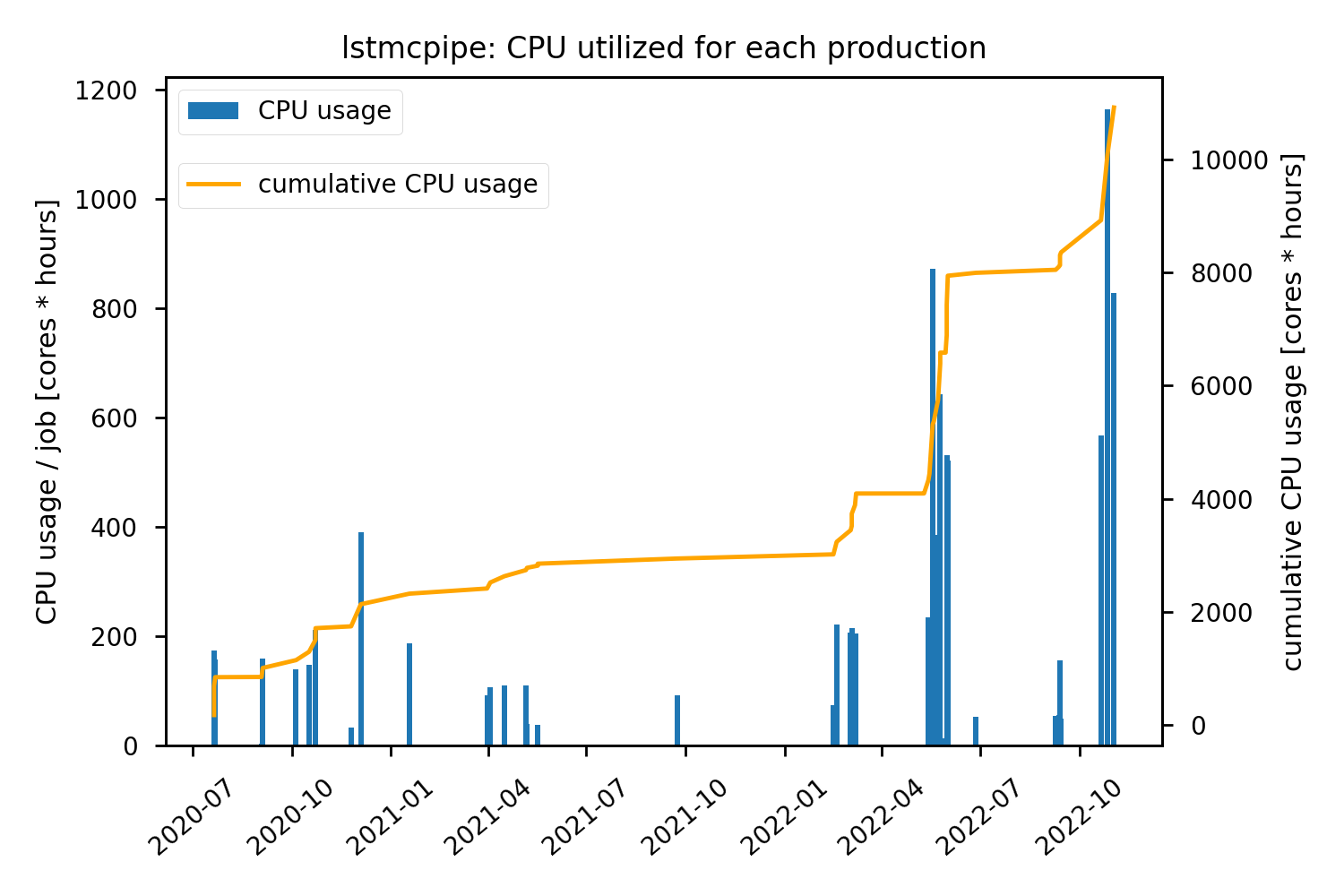}
    \end{minipage}%
    \begin{minipage}{0.5\textwidth}
        \centering
        \includegraphics[width=\textwidth]{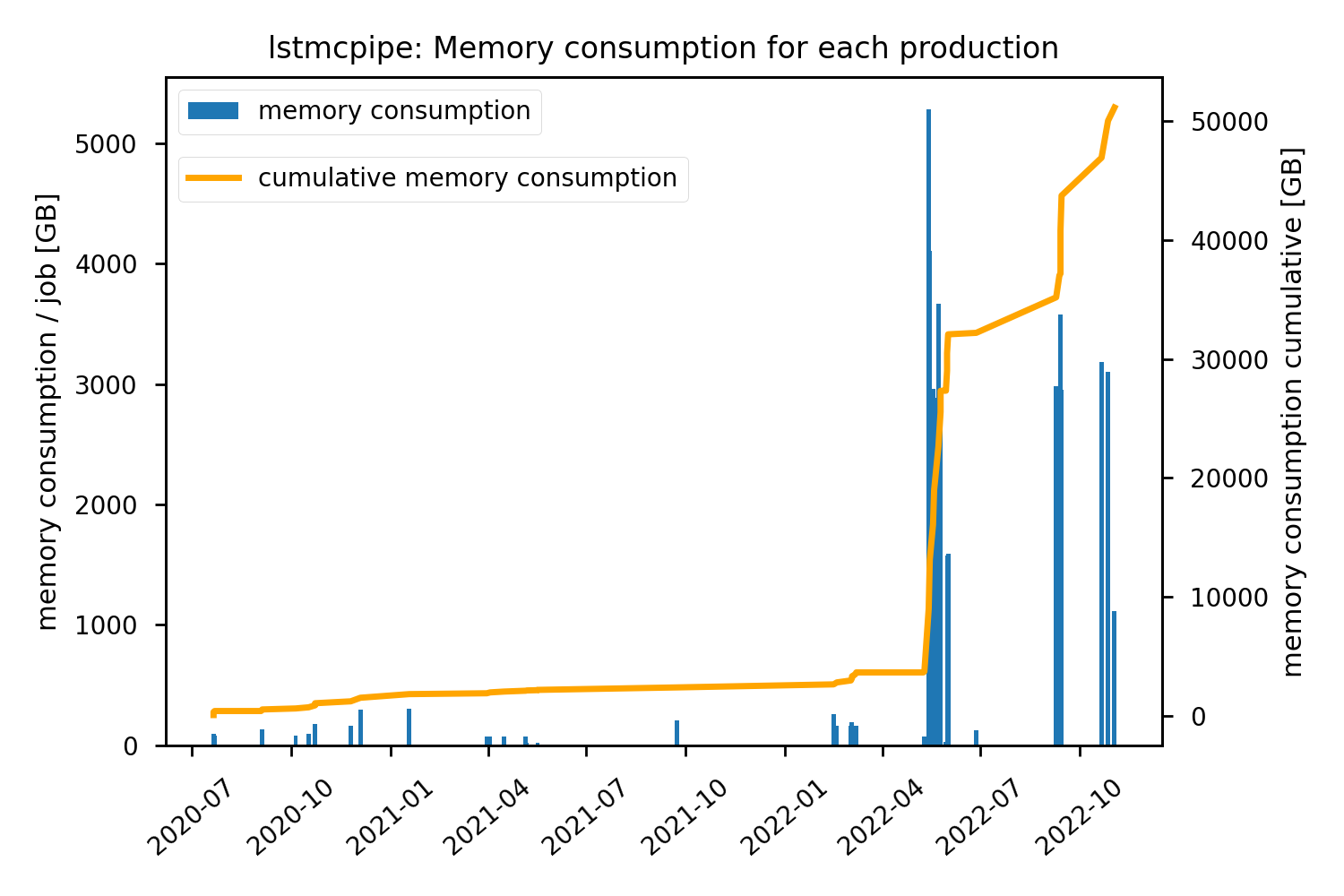}
    \end{minipage}
\caption{Computing resources used for the MC data analysis since July 2020. On the left the CPU usage per job and accumulated over time. On the right, the same for RAM usage.}
\label{fig:computing}
\end{figure}

Centralizing the MC data analysis also allows us to keep track more easily of the usage of computing resources required for their reduction for the LST analyzers (see Figure~\ref{fig:computing}). We notice the increase in the frequency of productions when we introduced the analysis as a service in April 2022. The increase in computing resources per job at a similar time is due to a newer and larger MC data set used for the analysis of LST-1 data.

\section{Conclusion}

We presented the \lstmcpipe library that is currently in production status and is used in the LST collaboration for the reduction of MC data, the training of random forest models and the production of IRFs needed for the data analysis. The library is open-source, can be installed as a PyPI package and its documentation can be found online.

\acknowledgements \small{The ASP would like to thank the dedicated researchers who are publishing with the ASP. ESCAPE - The European Science Cluster of Astronomy \& Particle Physics ESFRI Research Infrastructures has received funding from the European Union's Horizon 2020 research and innovation programme under Grant Agreement no. 824064.}

\bibliography{P79}  

\begin{thebibliography}{}
\expandafter\ifx\csname natexlab\endcsname\relax\def\natexlab#1{#1}\fi
\expandafter\ifx\csname url\endcsname\relax
  \def\url#1{\texttt{#1}}\fi
\expandafter\ifx\csname urlprefix\endcsname\relax\def\urlprefix{URL }\fi
\providecommand{\eprint}[2][]{\url{#2}}

\bibitem[{{Bernl{\"o}hr}(2008)}]{2008APh....30..149B}
{Bernl{\"o}hr}, K. 2008, Astroparticle Physics, 30, 149. \eprint{0808.2253}

\bibitem[{{CTA Consortium}(2018)}]{cta2018}
{CTA Consortium} 2018, Science with the Cherenkov Telescope Array ({World}
  {Scientific}). \urlprefix\url{https://doi.org/10.1142%2F10986}

\bibitem[{{Heck} et~al.(1998){Heck}, {Knapp}, {Capdevielle}, {Schatz}, \&
  {Thouw}}]{1998cmcc.book.....H}
{Heck}, D., {Knapp}, J., {Capdevielle}, J.~N., {Schatz}, G., \& {Thouw}, T.
  1998, {CORSIKA: a Monte Carlo code to simulate extensive air showers.} (.)

\bibitem[{{López-Coto} et~al.(2021){López-Coto}, {Baquero}, {Bernardos},
  {Cassol}, {Foffano}, {Garc\'ia}, {Gliwny}, {Iwamura}, {Jacquemont}, {Jouvin},
  {Kobayashi}, {Moralejo}, {Morcuende}, {Neise}, {Nozaki}, {N\"{o}the}, C.,
  {Renier}, {Saha}, {Sakurai}, {Sitarek}, \& {Takahashi}}]{lstchain22}
{López-Coto}, R., {Baquero}, A., {Bernardos}, M.~I., {Cassol}, F., {Foffano},
  L., {Garc\'ia}, E., {Gliwny}, P., {Iwamura}, Y., {Jacquemont}, M., {Jouvin},
  L., {Kobayashi}, Y., {Moralejo}, A., {Morcuende}, D., {Neise}, D., {Nozaki},
  S., {N\"{o}the}, M., C., P., {Renier}, Y., {Saha}, L., {Sakurai}, S.,
  {Sitarek}, J., \& {Takahashi}, M. 2021, in Astronomical Society of the
  Pacific Conference Series, edited by J.~E. {Ruiz}, F.~{Pierfedereci}, \&
  P.~{Teuben}, vol. 532 of Astronomical Society of the Pacific Conference
  Series, 369

\bibitem[{Nöthe et~al.(2021)Nöthe, Kosack, Nickel, \&
  Peresano}]{ctapipe-icrc-2021}
Nöthe, M., Kosack, K., Nickel, L., \& Peresano, M. 2021, in Proceedings, 37th
  International Cosmic Ray Conference, vol. 395

\bibitem[{{Ruiz} et~al.(2021){Ruiz}, {Morcuende}, {Saha}, {Baquero},
  {Contreras}, \& {Aguado}}]{lstosa}
{Ruiz}, J.~E., {Morcuende}, D., {Saha}, L., {Baquero}, A., {Contreras}, J.~L.,
  \& {Aguado}, I. 2021, in Astronomical Society of the Pacific Conference
  Series, edited by J.~E. {Ruiz}, F.~{Pierfedereci}, \& P.~{Teuben}, vol. 532
  of Astronomical Society of the Pacific Conference Series, 357.
  \eprint{2101.09690}

\bibitem[{Vuillaume et~al.(2022)Vuillaume, Garcia, \&
  Nickel}]{vuillaume_thomas_2022_7180216}
Vuillaume, T., Garcia, E., \& Nickel, L. 2022, lstmcpipe. {If you use this
  software, please cite it using Zenodo from
  https://doi.org/10.5281/zenodo.6460727},
  \urlprefix\url{https://doi.org/10.5281/zenodo.7180216}

\bibitem[{Yoo et~al.(2003)Yoo, Jette, \& Grondona}]{slurm}
Yoo, A.~B., Jette, M.~A., \& Grondona, M. 2003, in Job Scheduling Strategies
  for Parallel Processing, edited by D.~Feitelson, L.~Rudolph, \&
  U.~Schwiegelshohn (Berlin, Heidelberg: Springer Berlin Heidelberg), 44

\end{thebibliography}

\end{document}